# Contradictions in Improving Speed of Virus Scanning

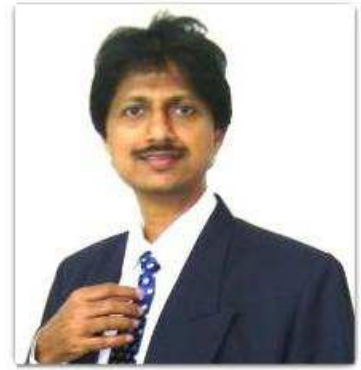

**By- Umakant Mishra, Bangalore, India**

umakant@trizsite.tk, http://umakant.trizsite.tk

**Contents**

## 1. Why the speed of virus scanning is a concern

According to Moore's law (http://en.wikipedia.org/wiki/Moore's_law) the CPU speed doubles in every two years. Not only the CPU speed, everything in computing industry grows in exponential rate including memory size, memory speed, storage space and others. But strangely there is no improvement in virus scanning time. Although the processing speed has substantially increased, a typical full scanning is still taking several hours for an average computer. There is a serious need to improve the scanning time.

Scanning speed is a serious concern because:

⇨ Slow scanning is very frustrating to users. Many users don't like to run anti-virus because it takes long time to scan.

⇨ fdsfsdf

⇨ In some situations like "boot time virus scan" we need fast scanning. The user cannot wait long time for the computer to boot.

⇨ "On access scanning" needs to scan the file fast before accessing the file. If the scanning takes long time the file access may give "time out" error.

⇨ Stream scanning and real time scanning has to be very fast in order to cope with the huge volume of data transfer in real time.

⇨ Network viruses spread very fast through Internet and create potential risk of viral outbreaks. Slow scanning methods are inefficient to tackle such fast spreading outbreaks.

## 2. Why does a virus scanning become slow

Scanning time is a function of various factors, such as, number of virus signatures, number of heuristics and other detection algorithms, number and size of files to be scanned and speed of the processor. A typical formula is as follows.

$$\text{Scanning time required} = \frac{\text{number of viruses X number and size of the file objects X number of detection methods}}{\text{Speed of processor}}$$

From the above formula it is pertinent that the scanning time can be reduced either by reducing parameters in the numerator, i.e., "number of files" etc. or by increasing the denominator, i.e., "speed of processor". As the speed of processor is increasing over time it should have naturally increased the scanning speed and reduced the scanning time. But why is there no improvement in scanning speed? That is because the factors in the denominator viz., number of files, number of viruses, complexity of viruses etc. are increasing even more rapidly than the processing speed. How to solve this problem?

Let's first see what are the factors that adversely affect the scanning speed and increase the scanning time.

⇨ Increase in the number and size of files

⇨ Increase in the number of viruses and virus signatures

⇨ Increase in the complexity of viruses

⇨ Increase in the number of scanning methods or algorithms (such as number of heuristics)

⇨ Need for a higher reliability in scanning

⇨ Increase in the complexity of compressed files- scanning compressed files needs more time to uncompress the files before scanning.

⇨ Repetitive scanning of the same files – either in the same computer or when transferred to other computers.

⇨ Storing and accessing previous scan data (to avoid repetitive scanning) also takes time

⇨ Weaker pattern matching logic can take more time

⇨ A weaker processor takes more time to scan

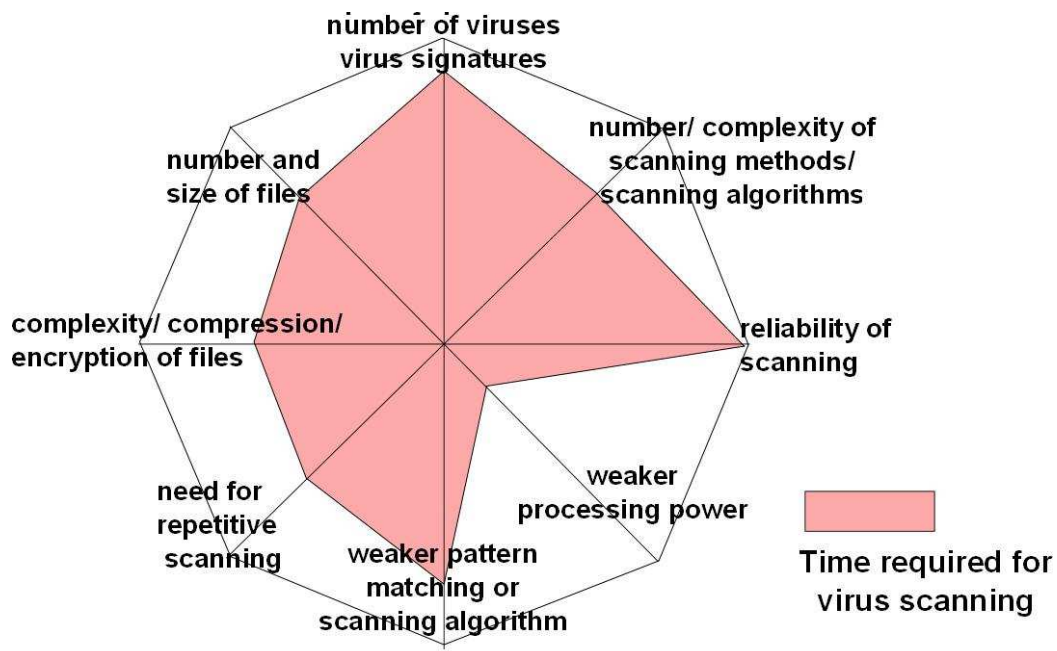

**Factors Affecting Scanning Speed**

It is pertinent from the above chart that there are certain factors like "reliability of scanning" which cannot be compromised. Similarly the factors like number and size of files, complexity of files, processing capability etc. which depend upon the user and his working environment where the anti-virus technology probably cannot do much. But there are factors like complexity of scanning methods, need for repetitive scanning etc. where the anti-virus technology can play a direct and significant role. Although the number of viruses will keep on increasing there can be methods to reduce the number of signatures (e.g., by using global signatures or by using generic methods of scanning). Thus the speed of virus scanning can be substantially increased by improving the anti-virus technology.

## 3. Contradiction of number of files and scanning time

The time required for scanning is directly correlated with the number and size of files in the computer. This is because all the data has to be read from the disk in order to be compared against virus signatures. An average processor takes about 30 minutes for scanning each 10 GB of hard disk. Thus if the number and size of the files are more then a scanning can take proportionately longer time because of the processing and disk I/O requirement.

| Contradiction | We want to scan all the files in order to ensure that no file is infected. But we don't want to scan all the files, as that would take long time. |
| | We want to increase the number of files and size of files in order to satisfy the growing need of our work. But we want to reduce the number of files (and file sizes) in order to have a faster scanning. |

## 4. Contradiction of number of virus signatures and scanning time

Anti-virus programs scan the computer files by comparing each file to a list of virus signatures. When the number of viruses increases, their signatures also increase and the required scanning time also increases as well.

| Contradiction | If a scanner includes all available virus signatures then the scanning process will take very long time. On the other hand if the scanner does not include some virus signatures then there is possibility of some viruses being escaped. |

# 5. Contradiction of more reliable scanning and scanning time

A scanning should be reliable and should have minimum number of false positives or false negatives. In order a scanning to be reliable it must do various tests to confirm the existence or non-existence of a virus. Thus reliable scanning involves various types of tests which requires more scanning time. Exact identification scanners use all signature strings of a virus, which are more reliable than nearly exactly identification scanners. But exact identification scanning takes more time to scan. The situation leads to following contradiction.

| Contradiction | If the anti-virus program does not apply all the required methods of scanning it may lead to false positives or false negatives. On the other hand if it applies all available methods to confirm a scan result, then it would take more scanning time. We want a confirmed result on scanning but we don't want to give more time for scanning. |
|---|---|

# 6. Contradiction of file security and scanning time

When a file is compressed or encrypted its bit pattern changes. As there can be numerous types of compression and encryption the scanner cannot compare the virus strings with the content of a compressed file. Hence, the scanner must decompress a compressed file and decrypt an encrypted file before scanning. This process of decompressing a file can take some time. But the situation becomes worse when the compression algorithm is too complex. The situation becomes even worse when the compressed/encrypted file is password protected.

| Contradiction | The anti-virus program cannot scan an encrypted file because its content is scrambled. People (and sometimes viruses) use complex compression/encryption algorithms for greater security and privacy but that situation requires complex scanning procedures and longer scanning time. |
|---|---|

# 7. Contradiction of Storing AV state data

In order to avoid repetitive scanning of the same files again and again in subsequent sessions the advanced scanners store the AV state information (or past scan information) in a database. The AV state information may include date and time of scanning, version of anti-virus update etc. The subsequent scanning sessions access this AV state database and avoid subsequent scanning of a previously scanned file.

But there are some problems involved in this method. The AV state data must be stored in a secured place which cannot be accessed by viruses. Besides storing of this data should be permanent in nature so as to be accessed by the anti-virus anytime in future. Thus storing and accessing the past scan information in a structured and secured database takes some time.

| Contradiction | If the past-scan information is stored in the cache then speed of accessing this data will be very fast. But the data in a cache is stored only for a limited period of time as the cache is volatile in nature. On the other hand, if the AV state is stored in a separate database, then accessing the database for storage and retrieval becomes slower. We want to store the AV state data permanently in a database but we don't want to make its access slow. |
|---|---|

# 8. How to solve contradictions of scanning speed

It is necessary to solve the above (and other) contradictions in order to increase the speed of scanning in different situations. Let's consider an example of "boot time scanning", a process that requires very fast scanning in order to avoid delay during the initial startup.

If we analyze a "boot time scanning" requirement we find that while it is necessary to scan the computer before booting, it is not necessary to scan all the files before booting. Hence the scanner can scan only the main memory, boot sector, system registry, operating system files and some other critical files (and objects) which might be enough to prevent an infection during a boot operation. As the scanning of a large number of files is avoided the scanning is finished within a very short time thereby without causing any delay in booting and without taking any risk of infection.

The above method involves the following techniques:

1. Identifying those files or system resources that are critical to be scanned before booting and differentiating from those that are not so critical to be scanned (Principle-2: Taking out).
2. Using advanced methods of scanning which can reduce scanning time (Principle-38: Improve Quality).
3. Skipping scanning of files which are not critical (Principle-21: Skipping)
4. Even the system files are not scanned for viral signatures, as that would require too much time. In order to save time they are often compared with a backup of their uninfected copies by an integrity checker which is a much faster method of detecting infection (Principle-26: Copying, Principle-28: Mechanics Substitution).

The above is an example resolving just one contradiction "increasing speed of boot time scanning without compromising security" by using TRIZ inventive principles. There are other methods of faster scanning such as "using wild card scanning", "advanced pattern matching", "using system cache for scanning", "scanning only the changed sectors", "using scanner cards", "using hardware based scanning", "using separate scan processor", "using virus co-processor", "scanning in a different machine", "smart scanning or intelligent scanning", "using faster databases", "using improved pattern matching", "using better methods for storing past scan data" etc. A proper application of all these methods can reduce the scanning time significantly.

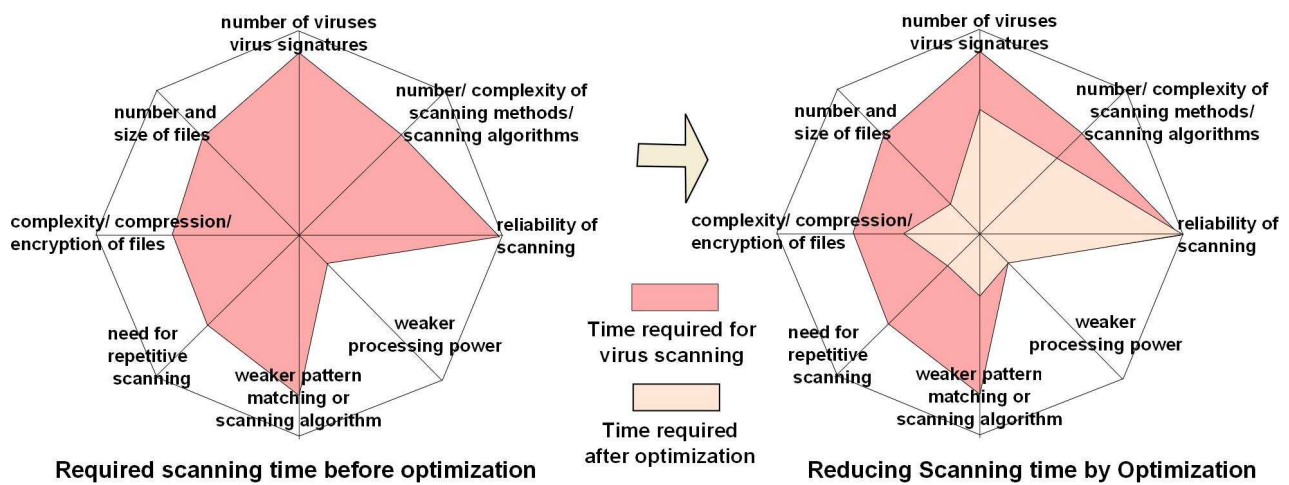

Required scanning time before optimization          Reducing Scanning time by Optimization

The above chart shows a comparative view of scanning time required before optimization and scanning time required after optimization. The right side chart shows the factors which can be optimized for scanning in order to reduce the required scanning time. Although the numbers of virus signature are increasing every day, there can be methods to group the viruses by their patterns and reduce the number of virus families and thereby reduce the number of required signatures.

Optimization of scanning is possible by solving the above contradictions and applying various other methods to improve the scanning technology. TRIZ provides powerful methods for solving contradictions, such as, Inventive Principles and Inventive Standards. However, the other methods of TRIZ like Ideality, Trends of Evolution, Resources and ARIZ can also be used to solve contradictions in some cases.

You may notice that the above graph shows a possibility of reducing the number of files. But how to reduce the number of files without affecting our working needs. Initially I also thought that the number of files couldn't be reduced for scanning. But then lately I realized that the number of files could be easily reduced. See below how.

## 9. Solving the contradiction of number of files

The number of files should increase (to store our programs, data and documents) but the number of files should decrease (for virus scanning). This physical contradiction is solved by TRIZ separation by condition. But practically how to reduce number of files for scanning without compromising security?

Actually we cannot reduce the total number of files in the user computer as those are created by the user or his operating environment. But we can reduce the number of files for scanning, for example, by avoiding scanning of certain file formats that does not support virus proliferation.

- Mark the previously scanned files and don't scan those file again. That way the number of files are reduced for subsequent scanning.

- Patent application 20080141375 (by Amundsen, June 2008) suggests to archive all the objects that are designated as non-recently used into a non-executable format. These archived objects are omitted from periodic virus scans. When there will be a request to use any archived object, the archived object will be converted to a non-archived state and virus scanned before made available to the requester. This way all archived objects are omitted in periodic virus scans.

We will discuss some more interesting example of solving contradictions in virus scanning in a separate paper.

# 10. Conclusion

Contradiction is a stage of problem solving where the nature of the actual problem is clearly explained in terms of at least two parameters, one improving and another worsening. While emphasizing one parameter strengthens the system position emphasizing another parameter weakens the system.

In conventional methods the problem solver has to make a perfect balance between these conflicting parameters, where the situation is neither too much on one side nor too much on the other. The results of those methods, although increase the speed of virus scanning, results in disadvantages like load on processor, increase in false positives and compromise on security. The objective of TRIZ is not to accept a tradeoff between the speed of scanning and those other difficulties but to resolve the contradictions so that the speed of scanning increases without compromising with security and other harmful results.